\documentstyle[12pt]{article}

\def\la{\langle}
\def\ra{\rangle}
\def\lb{\lbrack}
\def\rb{\rbrack}

\parskip=\medskipamount

\begin{document}

\makeatletter
\def\chapter{\@startsection{chapter}{1}{\z@}{-3.5ex plus -1ex minus -.2ex}
{2.3ex plus .2ex}{\Large\bf}}
\makeatother

\makeatletter
\def\section{\@startsection{section}{1}{\z@}{-3.5ex plus -1ex minus -.2ex}
{2.3ex plus .2ex}{\large\bf}}
\makeatother

\makeatletter
\@addtoreset{equation}{section}
\renewcommand{\theequation}{\thesection.\arabic{equation}}
\makeatother

\vspace{8mm}

\begin{center}

{\Large \bf The semiclassical approximation for the Chern--Simons partition
function } \\

\vspace{12mm}

{\large David H. Adams }

\vspace{4mm}

School of Mathematics, Trinity College, Dublin 2, Ireland. \\

\vspace{1ex}

email: dadams@maths.tcd.ie

\vspace{1ex}

19 September, 1997

\end{center}

\begin{abstract}

The semiclassical approximation for the partition function in Chern--Simons
gauge theory is derived using the invariant integration method. 
Volume and scale factors which were undetermined and had to be fixed by hand
in previous derivations are automatically taken account of in this framework.
Agreement with Witten's exact expressions for the partition function in the
weak coupling (large $k$) limit is verified for gauge group $SU(2)$ and 
spacetimes $S^3\,$, $S^2\times{}S^1\,$, $S^1\times{}S^1\times{}S^1$ 
and $L(p,q)$.

\end{abstract}

\newpage

\section{Introduction}

There has been much interest in the pure Chern--Simons gauge theory with 
spacetime a general 3-dimensional manifold since E.~Witten in 1989 gave
a prescription for obtaining exact expressions for the partition function
and expectation values of Wilson loops \cite{W(Jones)}. This prescription,
which is based on a correspondence with 2D conformal field theory, leads in
the case of the partition function to a new topological invariant of the
3-manifold\footnote{This was subsequently shown by K.~Walker \cite{Walker}
to coincide with the 3-manifold invariant constructed in a rigorous framework
by Reshetikhin and Turaev \cite{RT}.}, and in the case of Wilson loops
it leads to the Jones knot polynomial (and generalisations).
However, it is far from clear that Witten's exact prescription is compatible
with standard approaches to quantum field theory, in particular with 
perturbation theory. It is of great theoretical interest to compare results
obtained by Witten's prescription with those obtained by standard approaches;
this may lead to new insights into the scope or limitations of quantum field 
theory in general.

A basic prediction of perturbation theory is that the partition function
should coincide with its semiclassical approximation in the weak coupling
limit, corresponding to the asymptotic limit of large $k$ in the case of
Chern--Simons gauge theory. This has been investigated in a program initiated
by D.~Freed and R.~Gompf \cite{FG}, and followed up by other authors
\cite{J,Roz,AdSe(PLB)}. In these works the large $k$ asymptotics of the
expressions for the partition function obtained from Witten's prescription
was evaluated for various classes of spacetime 3-manifolds with gauge 
group $SU(2)$ and compared with expressions for the semiclassical 
approximations. Agreement was obtained, but only after fixing by hand the
values of certain undetermined quantities (volume and scale factors) appearing
in the expressions for the semiclassical approximation.

Our aim in this paper is to derive a complete, self-contained expression
for the semiclassical approximation for the Chern--Simons partition function
in which undetermined quantities do not appear. We do this using the invariant
integration method introduced by A.~Schwarz in \cite[App. II]{Sch(Inst)}.
An important property of the resulting expression is that it is independent
of the choice of invariant inner product in the Lie algebra of $G$ which is 
required to evaluate it. We provide an explicit demonstration of this; 
it amounts to showing that the expression is independent 
of the scale parameter $\lambda$ determining the inner product
$\la{}a,b\ra=-\frac{1}{\lambda}\mbox{Tr}(ab)$ in the Lie algebra.

The resulting expression for the semiclassical approximation is explicitly
evaluated for gauge group $SU(2)$ and spacetime 3-manifolds 
$S^3\,$, $S^2\times{}S^1\,$, $S^1\times{}S^1\times{}S^1$ and arbitrary lens
space $L(p,q)$. The expression for the semiclassical approximation involves
an integral over the moduli space of flat gauge fields, and the calculations
involve cases where the moduli space is both discrete ($S^3\,$, $L(p,q)$)
and continuous ($S^2\times{}S^1\,$, $S^1\times{}S^1\times{}S^1$).
After including the standard geometric counterterm in the phase factor
we find complete agreement with the exact expressions for the partition 
function in the large $k$ limit. The techniques and mathematics involved
in Witten's exact prescription (conformal blocks/representation theory
for Kac--Moody algebras, surgery techniques) are very different from those
used to obtain the semiclassical approximation (gauge theory, Hodge theory,
analytic continuation of zeta- and eta-functions). It is remarkable
that not only the general features such as the asymptotic $k$-dependence,
but also the precise numerical factors (including factors of $\pi$), are
reproduced by the semiclassical approximation.
Some details of our calculations, along with a more detailed description of
the invariant integration method, will be provided in a forthcoming 
paper \cite{Ad(prep)}.

\section{The invariant integration method}

We briefly recall the invariant integration method of Schwarz 
\cite[App. II]{Sch(Inst)}. Let $M$ be a closed manifold, $G$ a compact simple
Lie group, ${\bf g}$ the Lie algebra of $G\,$, $\Omega^q(M,{\bf g})$ the
space of q-forms on $M$ with values in ${\bf g}\,$, 
${\cal A}=\Omega^1(M,{\bf g})$ the space of gauge fields (for simplicity we
are assuming trivial bundle structure although this is not necessary),
${\cal G}$ the group of gauge transformations, i.e. the maps
$\phi:M\to{}G$ acting on ${\cal A}$ by $\phi\cdot{}A=\phi{}A\phi^{-1}+
\phi{}d\phi^{-1}$. A choice of metric on $M$ and $G$-invariant inner product
in ${\bf g}$ determine a ${\cal G}$-invariant inner product in each 
$\Omega^q(M,{\bf g})$ and ${\cal G}$-invariant metrics on ${\cal A}$
and ${\cal G}\,$, which in turn determine a metric on ${\cal A}/{\cal G}$.
Let $d_q^A:\Omega^q(M,{\bf g})\to\Omega^{q+1}(M,{\bf g})$ denote the 
exterior derivative twisted by gauge field $A$ (then $-d_0^A=-\nabla^A$
is the generator of infinitesimal gauge transformations of $A$).

Consider the partition function of a gauge theory with action functional
$S(A)\,$, formally given by 
\begin{eqnarray}
Z(\alpha)=\frac{1}{V({\cal G})}\int_{\cal A}{\cal D\/}A\,
e^{-\frac{1}{\alpha^2}S(A)}
\label{1}
\end{eqnarray}
where $\alpha$ is the coupling parameter and $V({\cal G})$ is the formal 
volume of ${\cal G}$. Rewrite:
\begin{eqnarray}
Z(\alpha)=\frac{1}{V({\cal G})}\int_{{\cal A}/{\cal G}}{\cal D\/}\lb{}A\rb\,
V(\lb{}A\rb)e^{-\frac{1}{\alpha^2}S(A)}
\label{2}
\end{eqnarray}
where $\lb{}A\rb={\cal G}\cdot{}A\,$, the orbit of ${\cal G}$ through $A\,$,
and $V(\lb{}A\rb)$ is its formal volume. Let ${\cal C}\subset{\cal A}$
denote the subspace of absolute minima for $S\,$, and 
${\cal M}={\cal C}/{\cal G}$ its moduli space. For $A_{\theta}\in{\cal C}$
expand
\begin{eqnarray}
S(A_{\theta}+B)=S(A_{\theta})+S_{A_{\theta}}^{(2)}(B)+
S_{A_{\theta}}^{(3)}(B)+\dots
\label{3}
\end{eqnarray}
where $S_{A_{\theta}}^{(p)}(B)$ is of order $p$ in $B\in\Omega^1(M,{\bf g})$.
Then the asymptotics of $Z(\alpha)$ in the limit $\alpha\to0$ is given by
\begin{eqnarray}
Z_{sc}(\alpha)=\frac{1}{V({\cal G})}\int_{\cal M}{\cal D\/}\lb{}A_{\theta}\rb\,
V(\lb{}A_{\theta}\rb)e^{-\frac{1}{\alpha^2}S(A_{\theta})}
\int_{\widetilde{T}_{\lb{}A_{\theta}\rb}}
{\cal D\/}\lb{}B\rb\,e^{-\frac{1}{\alpha^2}S_{A_{\theta}}^{(2)}(B)}
\label{4}
\end{eqnarray}
where $\widetilde{T}_{\lb{}A_{\theta}\rb}=T_{\lb{}A_{\theta}\rb}
({\cal A}/{\cal G})\Big/T_{\lb{}A_{\theta}\rb}{\cal M}$. Writing
\begin{eqnarray}
S_{\lb{}A_{\theta}\rb}^{(2)}(B)=\la{}B\,,D_{A_{\theta}}B\ra
\label{5}
\end{eqnarray}
where $D_{A_{\theta}}$ is a uniquely determined self-adjoint operator
on $\Omega^1(M,{\bf g})$ with $\ker(D_{A_{\theta}})=T_{A_{\theta}}{\cal C}$
one gets 
\begin{eqnarray}
\int_{\widetilde{T}_{\lb{}A_{\theta}\rb}}
{\cal D\/}\lb{}B\rb\,e^{-\frac{1}{\alpha^2}S_{A_{\theta}}^{(2)}(B)}
&=&\int_{\ker(D_{A_{\theta}})^{\perp}}{\cal D\/}B\,e^{-\frac{1}{\alpha^2}
\la{}B,D_{A_{\theta}}B\ra} \nonumber \\
&=& \det{}'\Bigl(\frac{1}{\pi\alpha^2}D_{A_{\theta}}\Bigr)^{-1/2}
\label{6}
\end{eqnarray}
Let $H_{A_{\theta}}\subset{\cal G}$ denote the isotropy subgroup of 
$A_{\theta}\,$, i.e. the subgroup of gauge transformations which leave
$A_{\theta}$ invariant. $H_{A_{\theta}}$ consists of constant gauge 
transformations, corresponding to a subgroup of $G$ which we also denote
by $H_{A_{\theta}}$ (see, e.g., \cite[p.132]{Donaldson}). Using the one-to-one
map
\begin{eqnarray*}
{\cal G}/_{H_{A_{\theta}}}\stackrel{\cong}{\longrightarrow}
{\cal G}\cdot{}A_{\theta}=\lb{}A_{\theta}\rb\qquad,\qquad\ \phi\mapsto
\phi\cdot{}A_{\theta}
\end{eqnarray*}
one gets 
\begin{eqnarray}
V(\lb{}A_{\theta}\rb)=|\det{}'(d_0^{A_{\theta}})|V({\cal G}/_{H_{A_{\theta}}})
=V({\cal G})V_{\cal G}(H_{A_{\theta}})^{-1}\det{}'((d_0^{A_{\theta}})^*
d_0^{A_{\theta}})^{1/2}
\label{7}
\end{eqnarray}
where $V_{\cal G}(H_{A_{\theta}})$ is the volume of $H_{A_{\theta}}$ 
considered as a subspace of ${\cal G}$. Substituting (\ref{6}) and (\ref{7})
in (\ref{4}) leads to Schwarz's expression for the semiclassical approximation
\cite[App.II, eq.(9)]{Sch(Inst)}:
\begin{eqnarray}
Z_{sc}(\alpha)=\int_{\cal M}{\cal D\/}\lb{}A_{\theta}\rb\,
V_{\cal G}(H_{A_{\theta}})^{-1}e^{-\frac{1}{\alpha^2}S(A_{\theta})}
\det{}'((d_0^{A_{\theta}})^*d_0^{A_{\theta}})^{1/2}
\det{}'\Bigl(\frac{1}{\pi\alpha^2}D_{A_{\theta}}\Bigr)^{-1/2}
\label{8}
\end{eqnarray}
In the cases of interest ${\cal M}$ is finite-dimensional (e.g. in the 
Yang--Mills theory it is an instanton moduli space), and the determinants in
(\ref{8}) can be zeta-regularised, leading to a finite expression
for $Z_{sc}(\alpha)$ (modulo any difficulties that may arise from ${\cal M}$
not being a smooth compact manifold).

\section{The semiclassical approximation in Chern--Simons gauge theory}

In Chern--Simons gauge theory, with 3-dimensional $M$ and 
gauge group $G=SU(N)\,$, the negative number
$-\frac{1}{\alpha^2}$ in (\ref{1}) is replaced by the purely imaginary 
number $ik$ ($k\in{\bf Z}$). It is therefore natural to take ${\cal C}$
to be the set of all critical points for the Chern--Simons action functional
in this case. Then the elements $A_{\theta}$ of ${\cal C}$ are the flat
gauge fields and ${\cal M}$ is the moduli space of flat gauge fields.
Expanding the Chern--Simons action functional
\begin{eqnarray}
S(A)=\frac{1}{4\pi}\int_M\mbox{Tr}(A\wedge{}dA+
{\textstyle \frac{2}{3}}A\wedge{}A\wedge{}A)
\label{9}
\end{eqnarray}
around a flat gauge field $A_{\theta}$ one finds
\begin{eqnarray}
S_{A_{\theta}}^{(2)}(B)=\frac{1}{4\pi}\int_M\mbox{Tr}(B\wedge{}
d_1^{A_{\theta}}B)
\label{10}
\end{eqnarray}
To obtain $D_{A_{\theta}}$ from this we need a metric on $M$ and invariant
inner product in ${\bf g}$ to determine the inner product in 
$\Omega^1(M,{\bf g})$. The $G$-invariant inner products in ${\bf g}$ are
those of the form 
\begin{eqnarray}
\la{}a,b\ra_{\bf g}=-\frac{1}{\lambda}\mbox{Tr}(ab)
\label{10a}
\end{eqnarray}
specified by the scale parameter $\lambda\in{\bf R}_+$. Thus:
\begin{eqnarray}
S_{A_{\theta}}^{(2)}(B)=\la{}B\,,D_{A_{\theta}}B\ra\qquad,\qquad\ 
D_{A_{\theta}}=-\frac{\lambda}{4\pi}\ast{}d_1^{A_{\theta}}
\label{11}
\end{eqnarray}
where $\ast$ is the Hodge operator. Using the regularisation procedure 
of \cite{AdSe(PLB)} we get
\begin{eqnarray}
\det{}'\Bigl(\frac{-ik}{\pi}D_{A_{\theta}}\Bigr)^{-1/2}
&=&\det{}'\Bigl(\frac{ik\lambda}{4\pi^2}\ast{}d_1^{A_{\theta}}\Bigr)^{-1/2}
\nonumber \\
&=&e^{\frac{i\pi}{4}\eta(A_{\theta})}\Bigl(\frac{k\lambda}{4\pi^2}
\Bigr)^{-\zeta(A_{\theta})/2}\det{}'((d_1^{A_{\theta}})^*
d_1^{A_{\theta}})^{-1/4}
\label{12}
\end{eqnarray}
where $\eta(A_{\theta})$ and $\zeta(A_{\theta})$ are the analytic continuations
to $s\!=\!0$ of the eta function $\eta(s;\ast{}d_1^{A_{\theta}})$ and
zeta function $\zeta(s;|\ast{}d_1^{A_{\theta}}|)$ respectively.
In \cite[eq. (23)]{AdSe(PLB)} we showed that
\begin{eqnarray}
\zeta(A_{\theta})=\dim{}H^0(A_{\theta})-\dim{}H^1(A_{\theta})
\label{13}
\end{eqnarray}
where $H^q(A_{\theta})$ is the q'th cohomology space of $d^{A_{\theta}}$.
Substituting in (\ref{8}) gives the following expression for the semiclassical
approximation for the Chern--Simons partition function:
\begin{eqnarray}
Z_{sc}(k)=\int_{\cal M}{\cal D\/}\lb{}A_{\theta}\rb\,
V_{\cal G}(H_{A_{\theta}})^{-1}e^{i(\frac{\pi}{4}\eta(A_{\theta})+
kS(A_{\theta}))}\Bigl(\frac{k\lambda}{4\pi^2}\Bigr)^{-\zeta(A_{\theta})/2}
\tau{}'(A_{\theta})^{1/2}
\label{14}
\end{eqnarray}
where $\tau{}'(A_{\theta})$ is the Ray--Singer torsion of $d^{A_{\theta}}$
\cite{RS(Adv)}. Note that this expression does not involve any undetermined 
quantities; all its ingredients are determined by the choice of metric on
$M$ and scale parameter $\lambda$ in the invariant inner product for ${\bf g}$.
We will discuss below the metric dependence of (and its removal from) this
expression. But first we derive the following:

\noindent {\it Theorem.} The semiclassical approximation for the Chern--Simons
partition function given by (\ref{14}) is independent of the scale parameter
$\lambda$.

{\it Proof.} Since a scaling of the inner product in ${\bf g}$ is equivalent 
to a scaling of the metric on $M$ in (\ref{14}) the theorem can be obtained 
by the general metric-independence arguments of Schwarz in 
\cite[\S5]{Sch(degen)}. But let us give an explicit derivation.
We will show that the $\lambda$-dependence of $V(H_{A_{\theta}})$
and ${\cal D\/}\lb{}A_{\theta}\rb$ factors out as 
\begin{eqnarray}
V(H_{A_{\theta}})&\sim&\lambda^{-(\dim{}H_{A_{\theta}})/2} \label{15a} \\
{\cal D\/}\lb{}A_{\theta}\rb&\sim&\lambda^{-(\dim{}T_{\lb{}A_{\theta}\rb}
{\cal M})/2} \label{16a}
\end{eqnarray}
Then, since $\dim{}H_{A_{\theta}}=\dim{}H^0(A_{\theta})$ and (in the generic
case) $\dim{}T_{\lb{}A_{\theta}\rb}{\cal M}=\dim{}H^1(A_{\theta})\,$,
we have
\begin{eqnarray}
{\cal D\/}\lb{}A_{\theta}\rb\,V(H_{A_{\theta}})^{-1}\;\sim\;
\lambda^{(\dim{}H^0(A_{\theta})-\dim{}H^1(A_{\theta}))/2}\,.
\label{17a}
\end{eqnarray}
This $\lambda$-dependence cancels against the $\lambda$-dependence of
$(k\lambda/4\pi^2)^{-\zeta(A_{\theta})/2}$ in (\ref{14}) due to (\ref{13}).
It is easy to see that all the other ingredients in (\ref{14}) are
independent of $\lambda$ and it follows that (\ref{14}) is 
$\lambda$-independent as claimed.

To derive (\ref{15a})--(\ref{16a}) we begin with a general observation on
the change in the volume element under a scaling of inner product in a
vectorspace $U$ of dimension $d$. Let $u_1,\dots,u_d$ be an orthonormal basis
for $U\,$, then the volume element $\mbox{vol}\in\Lambda^dU^*$ is the dual
of $u_1\wedge\cdots\wedge{}u_d\in\Lambda^dU\,$, i.e.  
$\mbox{vol}(u_1\wedge\cdots\wedge{}u_d)=1$. If we scale the inner product in
$U$ by $\la\cdot\,,\cdot\ra\to\la\cdot\,,\cdot\ra_{\lambda}=\frac{1}{\lambda}
\la\cdot\,,\cdot\ra$ then an orthonormal basis for the new inner product
is $u_1^{\lambda},\dots,u_d^{\lambda}$ where 
$u_j^{\lambda}=\sqrt{\lambda}\,u_j$. The new volume element 
$\mbox{vol}_{\lambda}\,$, given by $\mbox{vol}_{\lambda}(u_1^{\lambda}\wedge
\cdots\wedge{}u_d^{\lambda})=1\,$, is $\mbox{vol}_{\lambda}=
\lambda^{-d/2}\mbox{vol}$. The relations (\ref{15a}) and (\ref{16a}) follow
from this observation together with the fact that the metrics on 
$H_{A_{\theta}}$ and ${\cal M}$ depend on $\lambda$ through a factor 
$\frac{1}{\lambda}$ due to (\ref{10a}). This completes the 
proof.\footnote{In a previous preprint \cite{AdSe(hep-th)} we arrived at a
$\lambda$-dependent expression for the semiclassical approximation
for $M=S^3$. This was due to an error in our calculation equivalent to
assuming $\mbox{vol}_{\lambda}=\lambda^{d/2}\mbox{vol}$ instead of
$\lambda^{-d/2}\mbox{vol}$ in the argument above.}

Were it not for the metric-dependent phase factor 
$e^{\frac{i\pi}{4}\eta(A_{\theta})}$ the semiclassical approximation 
(\ref{14}) would be independent of the metric on $M$ by the general arguments
in \cite[\S5]{Sch(degen)}. A related observation is that, as it
stands, (\ref{14}) cannot reproduce Witten's exact formulae for the 
partition function at large $k$ because the latter are not only 
metric-independent but also involve a choice of framing of $M$. 
In \cite{W(Jones)} Witten resolved both of these problems by putting in by 
hand in the semiclassical approximation
a phase factor (``geometric counterterm'') depending both on the metric
and on the framing of $M$. It cancels the metric-dependence of 
$e^{\frac{i\pi}{4}\eta(A_{\theta})}$ and transforms under a change of 
framing in the same way as the exact expression for the partition
function. We will also put in this factor here.
We will carry out the calculations in the canonical
framing of Atiyah \cite{Atiyah(Top)}; then the inclusion of the geometric
counterterm in the phase amounts to replacing 
$\eta(A_{\theta})\to\eta(A_{\theta})-\eta(0)$ in (\ref{14}) \cite{FG}.

To explicitly evaluate (\ref{14}) we use the fact that the moduli space
${\cal M}$ can be identified with $\mbox{Hom}(\pi_1(M),G)/G\,$, the space
of homomorphisms $\pi_1(M)\to{}G$ modulo the conjugation action of $G$.
This leads, at least for the examples we consider below, to a one-to-one
correspondence of the form
\begin{eqnarray}
\widetilde{{\cal M}}\,\equiv\,(G_1\times\cdots\times{}G_s)/_{W}
\;\stackrel{\cong}{\longleftrightarrow}\;{\cal M}\qquad,\qquad\ 
\theta\leftrightarrow\lb{}A_{\theta}\rb
\label{18a}
\end{eqnarray}
where each $G_i$ is a subspace of $G\,$, $W$ is a finite group acting on the
$G_i$'s and $s$ is the number of generators of $\pi_1(M)$ which can be
independently associated with elements of $G$ to determine a homomorphism
$\pi_1(M)\to{}G$.
The inner product (\ref{10a}) determines a measure ${\cal D\/}\theta$
on $\widetilde{{\cal M}}\,$, and 
\begin{eqnarray}
{\cal D\/}\lb{}A_{\theta}\rb=|J_1(\theta)|{\cal D\/}\theta
\label{20a}
\end{eqnarray}
where the Jacobi determinant $|J_1(\theta)|$ depends only on the metric
on $M$ and ${\cal D\/}\theta$ depends only on $\lambda$. 
Similarly,
\begin{eqnarray}
V_{\cal G}(H_{A_{\theta}})=|J_0(\theta)|V(H_{A_{\theta}})
\label{21a}
\end{eqnarray}
where $V(H_{A_{\theta}})$ is the volume of $H_{A_{\theta}}$ as a subspace
of $G\,$, depending only on $\lambda\,$, and $|J_0(\theta)|$ depends only
on the metric on $M$. (Explicitly, $|J_0(\theta)|=
V(M)^{(\dim{}H_{A_{\theta}})/2}$.) 
Putting all this into (\ref{14}) gives
\begin{eqnarray}
Z_{sc}(k)=\int_{\widetilde{\cal M}}{\cal D\/}\theta\,
V(H_{A_{\theta}})^{-1}e^{i\lb\frac{\pi}{4}(\eta(A_{\theta})-\eta(0))+
kS(A_{\theta})\rb}\Bigl(\frac{k\lambda}{4\pi^2}\Bigr)^{-\zeta(A_{\theta})/2}
\tau(A_{\theta})^{1/2}
\label{15}
\end{eqnarray}
where 
\begin{eqnarray}
\tau(A_{\theta})^{1/2}=|J_0(\theta)|^{-1}|J_1(\theta)|
\tau{}'(A_{\theta})^{1/2}\,.
\label{16}
\end{eqnarray}
This quantity is the square root of the Ray--Singer torsion ``as a function
of the cohomology'', introduced and shown to be metric-independent in
\cite[\S3]{RS(AMS)}. Since $\eta(A_{\theta})-\eta(0)$ is known to be
metric-independent \cite{APS(II)} we see that the resulting expression
(\ref{15}) for $Z_{sc}(k)$ is metric-independent as discussed above.

\section{Explicit evaluations of the semiclassical approximation}

We evaluate $Z_{sc}(k)$ in the cases where $G=SU(2)$ and
$M$ is $S^3\,$, $S^2\times{}S^1\,$, $S^1\times{}S^1\times{}S^1$ and
$L(p,q)\,$, and compare with the expressions $Z_W(k)$ for the partition
function obtained from Witten's exact prescription in the large $k$ limit.
To do the calculations we must choose a value for $\lambda\,$;
the answers are independent of the choice due to the theorem in \S3.
A basis for ${\bf g}=su(2)$ is
\begin{eqnarray}
a_1={\textstyle\frac{1}{2}\left({0 \atop i}\;{ i \atop 0} \right)}\quad\quad
a_2={\textstyle\frac{1}{2}\left({0 \atop -1}\;{ 1 \atop 0} \right)}\quad\quad
a_3={\textstyle\frac{1}{2}\left({i \atop 0}\;{ 0 \atop -i} \right)}
\label{17}
\end{eqnarray}
Since $\mbox{Tr}(a_ia_j)=-\frac{1}{2}\delta_{ij}$ a convenient choice
for $\lambda$ is
\begin{eqnarray}
\lambda=1/2\,,
\label{18}
\end{eqnarray}
then $\{a_1,a_2,a_3\}$ is an orthonormal basis for $su(2)\,$, determining
a left invariant metric on $SU(2)$. The volume of $SU(2)$ corresponding to 
this metric can be calculated to be
\begin{eqnarray}
V(SU(2))=16\pi^2\,.
\label{19}
\end{eqnarray}
Define $U(1)\subset{}SU(2)$ by
\begin{eqnarray}
U(1)=\Big\{e^{a_3\theta}={\textstyle
\left({ e^{i\frac{\theta}{2}} \atop 0}\;{0 \atop 
e^{-i\frac{\theta}{2}}} \right)}\ \Big|\ \theta\in\lb{}0,4\pi\lb
\ \Big\}\ 
\label{20}
\end{eqnarray}
Since $a_3$ is a unit vector in $su(2)\,$, $\,\frac{d}{ds}\Big|_{s=0}
e^{a_3(\theta+s)}=a_3$ is a unit tangent vector to $SU(2)$ at $e^{a_3\theta}$
and it follows that the volume of $U(1)$ in $SU(2)$ is
\begin{eqnarray}
V(U(1))=\int_0^{4\pi}d\theta=4\pi\,.
\label{21}
\end{eqnarray}

\noindent $\underline{M=S^3}$:

$\pi_1(S^3)$ is trivial, so ${\cal M}$ consists of a single point 
corresponding to $A_{\theta}=0$.
Then $H_{A_{\theta}}=H_0=G=SU(2)\,$, $\,\dim{}H^0(0)\!=\!3\,$, 
$\dim{}H^1(0)\!=\!0\,$, $\zeta(0)=3-0=3$.
In \cite{Ad(prep)} we calculate the torsion (\ref{16}) to be
\begin{eqnarray}
\tau(0)=1
\label{21b}
\end{eqnarray}
Substituting in (\ref{15}) we get
\begin{eqnarray}
Z_{sc}(k)=\frac{1}{V(G)}\Bigl(\frac{k\lambda}{4\pi^2}\Bigr)^{-\zeta(0)/2}
\tau(0)^{1/2}
=\frac{1}{16\pi^2}\Bigl(\frac{k\frac{1}{2}}{4\pi^2}\Bigr)^{-3/2}
=\sqrt{2}\pi{}k^{-3/2}\,.
\label{22}
\end{eqnarray}
This coincides with the exact formula \cite[eq (2.26)]{W(Jones)} 
in the large $k$ limit:
\begin{eqnarray}
Z_W(k)=\sqrt{\frac{2}{k+2}}\sin\Bigl(\frac{\pi}{k+2}\Bigr)\ \;\sim\ \;
\sqrt{2}\pi{}k^{-3/2}\qquad\mbox{for $k\to\infty$}\,.
\label{23}
\end{eqnarray}

\noindent $\underline{M=S^2\times{}S^1}$:

$\pi_1(S^2\times{}S^1)\cong{\bf Z}\,$, so by standard arguments
\begin{eqnarray}
{\cal M}\;\cong\;\mbox{Hom}({\bf Z},SU(2))/_{SU(2)}\;\cong\;
U(1)/_{{\bf Z}_2}\,\equiv\,\widetilde{{\cal M}}
\label{24}
\end{eqnarray}
where $U(1)$ is given by (\ref{20}) and the action of ${\bf Z}_2$ on
$U(1)$ is generated by $e^{a_3\theta}\to{}e^{-a_3\theta}$. 
It follows that $\widetilde{\cal M}$ can be identified with
$\lb0,2\pi\rb$ and $\int_{\widetilde{{\cal M}}}{\cal D\/}\theta
(\cdots)=\int_{\lb0,2\pi\rb}d\theta(\cdots)$ where $\theta$ is the parameter 
in (\ref{20}). The isotropy group $H_{A_{\theta}}$ is the maximal 
subgroup of $SU(2)$ which commutes with $e^{a_3\theta}\,$, so 
$H_{A_{\theta}}=U(1)$ for $\theta\ne0$. Hence $V(H_{A_{\theta}})\!=\!4\pi\,$,
$\dim{}H^0(A_{\theta})\!=\!\dim{}H^1(A_{\theta})\!=\!1\,$, 
$\zeta(A_{\theta})\!=\!1\!-\!1\!=\!0$. 
In \cite{Ad(prep)} we show that
$S(A_{\theta})\!=\!0\,$, $\eta(A_{\theta})\!=\!0$ for all $A_{\theta}\,$,
and calculate 
\begin{eqnarray}
\tau(A_{\theta})=(2-2\cos\theta)^2\,.
\label{25}
\end{eqnarray}
Putting all this into (\ref{15}) we get
\begin{eqnarray}
Z_{sc}(k)&=&\int_{\lb0,2\pi\rb}d\theta\,\frac{1}{V(H_{A_{\theta}})}
\Bigl(\frac{k\lambda}{4\pi^2}\Bigr)^{-\zeta(A_{\theta})/2}
\tau(A_{\theta})^{1/2} \nonumber \\
&=&\int_{\lb0,2\pi\rb}d\theta\,\frac{1}{4\pi}
\Bigl(\frac{k\frac{1}{2}}{4\pi^2}\Bigr)^0(2-2\cos\theta)=1
\label{26}
\end{eqnarray}
which coincides with the exact formula \cite[eq. (4.31)]{W(Jones)}:
\begin{eqnarray}
Z_W(k)=1
\label{27}
\end{eqnarray}

\noindent $\underline{M=S^1\times{}S^1\times{}S^1}$:

$\pi_1(S^1\times{}S^1\times{}S^1)={\bf Z}\times{\bf Z}\times{\bf Z}\,$, so
by standard arguments
\begin{eqnarray}
{\cal M}\;\cong\;\mbox{Hom}({\bf Z}\times{\bf Z}\times{\bf Z}\,,SU(2))/_{SU(2)}
\;\cong\;(U(1)\times{}U(1)\times{}U(1))/_{{\bf Z}_2}
\;\equiv\;\widetilde{{\cal M}}
\label{28}
\end{eqnarray}
Set $\theta=(\theta_1,\theta_2,\theta_3)$ where $\theta_1\,$, $\theta_2\,$,
$\theta_3$ are 3 copies of the parameter for $U(1)$ in (\ref{20});
it follows from (\ref{28}) that $\widetilde{\cal M}$ can be identified with
$\lb0,2\pi\rb\times\lb0,4\pi\lb\times\lb0,4\pi\lb$ and
\begin{eqnarray*}
\int_{\widetilde{\cal M}}{\cal D\/}\theta(\cdots)=
\int_{\lb0,2\pi\rb\times\lb0,4\pi\lb\times\lb0,4\pi\lb}d\theta_1d\theta_2
d\theta_3(\cdots)
\end{eqnarray*}
We have $\dim{}H^1(A_{\theta})\!=\!\dim{}T_{\theta}\widetilde{\cal M}\!=\!3\,$,
$H_{A_{\theta}}\!=\!U(1)$ (except for $\theta\!=\!(0,0,0)$ and several
other isolated points), so $\dim{}H^0(A_{\theta})\!=\!\dim(U(1))\!=\!1$
and $\zeta(A_{\theta})\!=\!1\!-\!3\!=\!-2$. 
In \cite{Ad(prep)} we show that
$S(A_{\theta})\!=\!0\,$, $\eta(A_{\theta})\!=\!0$ for all $A_{\theta}\,$,
and calculate 
\begin{eqnarray}
\tau(A_{\theta})=1\,.
\label{30}
\end{eqnarray}
Putting all this into (\ref{15}) we get
\begin{eqnarray}
Z_{sc}(k)&=&\int_{\lb0,2\pi\rb\times\lb0,4\pi\lb\times\lb0,4\pi\lb}
d\theta_1d\theta_2d\theta_3\,\frac{1}{V(H_{A_{\theta}})}
\Bigl(\frac{k\lambda}{4\pi^2}\Bigr)^{-\zeta(A_{\theta})/2}
\tau(A_{\theta})^{1/2} \nonumber \\
&=&\int_{\lb0,2\pi\rb\times\lb0,4\pi\lb\times\lb0,4\pi\lb}
d\theta_1d\theta_2d\theta_3\,\frac{1}{4\pi}
\Bigl(\frac{k\frac{1}{2}}{4\pi^2}\Bigr)^1\cdot1 \nonumber \\
&=&k
\label{31}
\end{eqnarray}
which coincides in the large $k$ limit with the exact formula
\cite[eq.(4.32)]{W(Jones)}:
\begin{eqnarray}
Z_W(k)=k+1\,.
\label{32}
\end{eqnarray}

\noindent $\underline{M=L(p,q)}$:

In this case the quantities of interest have been calculated in 
\cite{FG,J} and we quote the results. $L(p,q)=S^3/_{{\bf Z}_p}$ for a 
certain free action of ${\bf Z}_p$ on $S^3$ specified by $p$ and $q$
(which must be relatively prime), so $\pi_1(L(p,q))={\bf Z}_p$ and
\begin{eqnarray}
{\cal M}\;\cong\;\mbox{Hom}({\bf Z}_p,SU(2))/_{SU(2)}
\;\cong\;\{e^{a_3(4\pi{}n/p)}\;|\;0\le{}n\le{}p/2\}\;\equiv\;
\widetilde{\cal M}
\label{33}
\end{eqnarray}
(see \cite{FG}). Thus the moduli space is discrete, $\theta\to{}n\,$,
$A_{\theta}\to{}A_n\,$, $\int_{\widetilde{\cal M}}{\cal D\/}\theta(\cdots)\to
\sum\limits_{0\le{}n\le{}p/2}(\cdots)$ and $\dim{}H^1(A_n)\!=\!
\dim{\cal M}\!=\!0$. The isotropy group $H_{A_n}$ is the maximal subgroup
of $SU(2)$ whose elements commute with $e^{a_3(4\pi{}n/p)}\,$, so for
$0<n<p/2\,$ $H_{A_n}\!=\!U(1)\,$, $V(H_{A_n})\!=\!4\pi\,$, 
$\dim{}H^0(A_n)\!=\!\dim{}H_{A_n}\!=\!1$ and $\zeta(A_n)\!=\!1\!-\!0\!=\!1$.
For $n\!=\!0\,$, and for $n\!=\!p/2$ if $p/2$ is integer, 
$e^{a_3(4\pi{}n/p)}\!=\!\pm1\,$, so in this case $H_{A_n}\!=\!SU(2)\,$,
$\dim{}H_{A_n}\!=\!3$ and $\zeta(A_n)\!=\!3\!-\!0\!=\!3$. By (\ref{15}),
the $k$-dependence of the summand is $\,\sim\,k^{-\zeta(A_n)/2}$ and it 
follows that the terms corresponding to $n\!=\!0$ and $n\!=\!p/2$ (if integer)
do not contribute to the large $k$ asymptotics of $Z_{sc}(k)\,$, so we
discard these in the following; i.e. restrict to $0<n<p/2$.
In \cite[eq.(5.3) and prop.5.2]{J} it was shown that
\begin{eqnarray}
e^{i\frac{\pi}{4}(\eta(A_n)-\eta(0))+kS(A_n)}
=ie^{2\pi{}iq^*(k+2)n^2/p}
\label{34}
\end{eqnarray}
where $q^*q\!=\!1$ (mod $p$). The torsion $\tau(A_n)$ is obtained from the
calculations in \cite{FG} to be
\begin{eqnarray}
\tau(A_n)=\frac{16}{p}\sin^2\Bigl(\frac{2\pi{}n}{p}\Bigr)\sin^2\Bigl(
\frac{2\pi{}q^*n}{p}\Bigr)
\label{35}
\end{eqnarray}
It follows that the large $k$ asymptotics of (\ref{15}) in this case is
\begin{eqnarray}
Z_{sc}(k)&\stackrel{k\to\infty}{\simeq}&\sum_{0<n<p/2}\frac{1}{V(H_{A_n})}
e^{i\frac{\pi}{4}(\eta(A_n)-\eta(0))+kS(A_n)}
\Bigl(\frac{k\lambda}{4\pi^2}\Bigr)^{-\zeta(A_n)/2}
\tau(A_n)^{1/2} \nonumber \\
&=&\sum_{n=1}^{\lb\frac{p-1}{2}\rb}\frac{1}{4\pi}ie^{2\pi{}iq^*(k+2)n^2/p}
\Bigl(\frac{k\frac{1}{2}}{4\pi^2}\Bigr)^{-1/2}\frac{4}{\sqrt{p}}
\Big|\sin\Bigl(\frac{2\pi{}n}{p}\Bigr)\sin\Bigl(\frac{2\pi{}q^*n}{p}\Bigr)\Big|
\nonumber \\
&=&\frac{\sqrt{2}}{\sqrt{k}}\sum_{n=1}^p
ie^{2\pi{}iq^*(k+2)n^2/p}\frac{1}{\sqrt{p}}
\Big|\sin\Bigl(\frac{2\pi{}n}{p}\Bigr)\sin\Bigl(\frac{2\pi{}q^*n}{p}
\Bigr)\Big|\,.
\label{36}
\end{eqnarray}
This is precisely the formula for the large $k$ asymptotics of the 
exact partition function $Z_W(k)$ derived in \cite[eq.(5.7)]{J}.

The calculation of $\tau(A_{\theta})$ in the preceding examples is carried
out in \cite{Ad(prep)} using the equality between Ray--Singer torsion and the
combinatorial R-torsion (both considered as functions of the cohomology)
\cite{Cheeger-Muller}. The R-torsion is calculated in each case using a 
suitable cell decomposition of $M$ for which the combinatorial objects 
corresponding to the $|J_q(\theta)|$'s in (\ref{16}) are equal to 1;
then the combinatorial object corresponding to $\tau{}'(A_{\theta})$
is calculated to give (\ref{21b}), (\ref{25}), (\ref{30}) and (\ref{35})
respectively. For the cases where $M$ is $S^2\times{}S^1$ and 
$S^1\times{}S^1\times{}S^1$ the result $\eta(A_{\theta})\!=\!0$ is obtained  
in \cite{Ad(prep)} by
decomposing $\Omega^1(M,{\bf g})$ into the direct sum of finite-dimensional
subspaces invariant under $\ast{}d_1^{A_{\theta}}\,$, and 
showing that $\ast{}d_1^{A_{\theta}}$ has symmetric spectrum on each of these 
subspaces.

In future work we will be checking the agreement between the expression
(\ref{15}) for the semiclassical approximation and the exact formulae
for the partition function in the large $k$ limit for other more complicated
situations, e.g. when $M$ is a torus bundle \cite{J,Andersen} or a Seifert
manifold \cite{Roz}, and for gauge groups other than $SU(2)$.
In certain cases, e.g. for certain Seifert manifolds, the $k$-dependence
$\,\sim\,k^{\stackrel{max}{\theta}\Big\{-\zeta(A_{\theta})/2\Big\}}$
predicted by the semiclassical approximation fails \cite{Roz}. This is
related to the moduli space ${\cal M}$ having certain singularities.
It is an interesting problem to refine the derivation of the semiclassical
approximation given here so that it also works for these cases.

{\bf Acknowledgements.} I am grateful to Siddhartha Sen for many helpful
discussions and encouragement during this work, which was supported by
Forbairt and Hitachi Labs, Dublin.

\end{document}